\documentclass[aps,preprint]{revtex4}
\input{epsf}
\usepackage[dvips]{graphicx}

\def\be{\begin{equation}}
\def\ee{\end{equation}}
\def\bea{\begin{eqnarray}}
\def\eea{\end{eqnarray}}
\def\ba{\begin{array}}
\def\ea{\end{array}}

\def\part{\partial}

\preprint{SUGP-02/2-3} \preprint{hep-th/0202160} \vspace{5cm}

\begin{document}

\title{Stabilization of Hyperbolic Brane-World Scenarios.}

\author{Pedro J. Silva}

\affiliation{~\\
Physics Department, \\
Syracuse University, \\
Syracuse, New York 13244-1130.}

\begin{abstract}
In this talk we consider the issue of stabilization of compact
hyperbolic brane-world scenarios from the point of view of
4-dimensional effective theories. The idea is to clarify the
status of stabilization for these models. Possible ways to
overcome a no-go theorem that appeared in a recent paper are shown
invoking the holographic framework and type IIA*/IIB* theories. A
brief discussion on supersymmetry is also given.
\end{abstract}

\maketitle

\section{Introduction}

In the last couple of years brane-world scenarios have received
much attention. These constructions are effective models, where
our 4-dimensional world is realized as a brane embedded in a
bigger universe, usually called the bulk space-time. Ideally this
picture should result from a low energy limit of a more
fundamental theory like string theory. In this short note we
address the problem of stabilization of the extra-dimensions for
the particular case of compact hyperbolic brane-world scenarios
(CHBS). In CHBS the bulk space-time is of the form $M_4 \otimes
H_d/\Gamma$,
 where $M_4$ is identified with our 4-dimensional world, $H_d$ is a
 d-dimensional hyperbolic manifold (a homogenous Euclidean space
with constant negative curvature) and $\Gamma$ is a discrete
subgroup of SO(d,1) acting freely. These types of constructions
enjoy many appealing characteristics that makes the investigation
of the stability issue worthwhile. For information on the
phenomenological implications of these models we refer the
original articles \cite{kmst, sst1,sst2}.

The main purpose of this talk is to present a detailed analysis of
radion stabilization in CHBS (see the original paper \cite{nsst}).
It has recently been demonstrated \cite{Carroll:2001ih} that, in
the context of general relativity in $4+d$ dimensions,
stabilization of large hyperbolic extra dimensions, leaving
Minkowski space on our brane, requires a violation of the null
dominant energy condition. Here we extend this argument to the
case in which our brane is allowed to exhibit standard FRW
expansion and comment on the regime of validity of this result. We
then turn to possible ways in which stabilization may work due to
a breakdown of the assumptions in the previous argument by
considering holography and type IIA*/IIB* supergravities.

In the second section, we show the flat brane case and recover the
results of \cite{Carroll:2001ih}. In the third section, we
generalize to FRW branes and in the fourth section, the inclusion
of extra compact directions like a n-sphere is considered. The
fifth section contains some comments on possible ways to bypass
the no-go-theorem and a small discussion on supersymmetry.

\section{Flat Brane-world and d-dimensional Hyperbolic manifold}

Our starting point is for the Einstein gravity in 4+d-dimensional
space-time, with bulk matter. \bea S_{d+4}=\int
dX^{4+d}\sqrt{-G}\left(M^{d+2}R(G)-{\cal L}_{bulk}\right), \eea
where $M$ is the $4+d$ plank mass, ${\cal L}_{bulk}$ stands for
the bulk matter field, the geometry is described by the metric
$ds^2=G_{AB}dX^AdX^B=\bar{g}_{\mu\nu}dx^\mu dx^\nu +
r^2\gamma_{ij}dx^idx^j$, where capital Latin letters runs over all
of the space-time dimensions, Greeks letters over the
4-dimensional brane-world and lower case Latin over the hyperbolic
manifold. $\bar{g}$ is the brane metric, $\gamma$ is the
hyperbolic metric of radius $d(d+1)$ and $r$ is the radion, that
we want to stabilized at the value $R_h$. Using that we are in the
case of compact hyperbolic manifold with volume $e^\alpha$, we
define the 4-dimensional planck mass $M^2_4=M^{2+d}R_h^de^\alpha$
and the field $\phi$ by the equation $r=R_h
e^{\sqrt{1/d(d+2)}\phi/M_4}$. Also we need a conformal rescaling
on the brane metric
$\bar{g}_{\mu\nu}=g_{\mu\nu}e^{-\sqrt{d/(d+2)}\phi/M_4}$, to
decouple the new field $\phi$ and the reduced Einstein tensor.
After some algebra we get the 4-dimensional effective action, \bea
S_{d+4}=&&\int dx^d{\sqrt{\gamma}\over e^\alpha}S_{eff}\;,\nonumber \\
S_{eff}=&&\int dx^{4}\sqrt{-g}\left[ M^2_4R(g)-{1\over
2}(\nabla\phi)^2 -W(\phi,g)\right],
\eea
where
\be W(\phi,g)=
\frac{M_4^2}{M^{d+2}}e^{-\sqrt{d/(d+2)}\phi/M_4}{\cal
L}_{bulk}+\frac{d(d-1)M_4^2}{R_h^2}e^{-\sqrt{(d+2)/d}\phi/M_4}
\ee
and we have used that $R(\gamma)=-d(d-1)$. The stabilization of
the radion translates into the following system of equations
\bea
&\partial_\phi W|_{\phi=0}=0\;\;\;,\;\;\;\partial^2_\phi
W|_{\phi=0}>0& \nonumber \\
&\left(g_{\mu\nu}W-2{\partial W \over \partial g^{\mu\nu}}
\right)|_{\phi=0}=0.& \label{eqn1} \eea To obtain information
about the energy conditions that the bulk matter has to obey in
other to satisfy this equations, we have to rewrite these
effective equations in terms of the stress energy tensor of the
bulk space-time. Here we use that the radion filed is really part
of the metric, therefore its field equations involve some
combinations of the stress energy tensor of the bulk matter
fields, i.e. \be \frac{\partial }{\partial\phi}= \frac{\partial
G^{\mu\nu}}{\partial\phi}\frac{\partial }{\partial
G^{\mu\nu}}+\frac{\partial G^{ij }}{\partial\phi}\frac{\partial
}{\partial G^{ij}}, \ee taking into account the conformal
transformation on the brane metric, part of the set of equations
\ref{eqn1} translates into \bea
&T_{\mu\nu}=\frac{d(d-1)M^{d+2}}{2R^2_H}G_{\mu\nu},& \nonumber \\
&dG^{\mu\nu}T_{\mu\nu}-2G^{ij}T_{ij}=\frac{d(d+2)(d-1)M^{2+d}}{R_h^2},&
\eea where we have assume that the stabilization is archive by the
matter field on the bulk. Next, using the following form for the
stress energy tensor, \bea
T_{tt}&=&\rho, \nonumber \\
T_{\alpha\beta}&=&p\,\Sigma_{\alpha\beta}, \nonumber\\
T_{ij}&=&qR^2_h\gamma_{ij}, \label{eqn2} \eea where $\mu$ has been
divided into time and the three space-like directions
$(t,\alpha)$. Then we get the constraints
\bea &&\rho=-\frac{d(d-1)M^{2+d}}{2R^2_h}, \nonumber\\
&&p=\frac{d(d-1)M^{2+d}}{2R^2_h},\nonumber\\
&&q=\frac{(d-1)(d-2)M^{2+d}}{4R^2_h}. \label{eqn3} \eea This type
of matter field does not obey the null energy condition as can be
easily seen by considering a general null-vector along the
hyperboloid, like $\textsl{l}=\partial_t+e_i$, where $e_i$ are
orthogonal vector basis on the hyperboloid. Then, contracting
$\textsl{l}$ twice with the stress energy tensor $T_{AB}$ gives
$-\frac{(d-1)M^{2+d}}{R^2_h}$, a negative number violating the
above energy condition. Therefore we conclude this part of the
discussion saying that: \vspace{.5cm}

\textit{Although in principle equation \ref{eqn3} can be
satisfied, the matter field required will not satisfy the null
energy condition.}

\vspace{.5cm} To clarify the above general statement, let us
consider a simple example of stabilization due to matter field
violating the null energy condition. Consider a flat brane-world
and as bulk matter a cosmological constant
$\Lambda=M^{d+4}\lambda$, also include a d-form over the
hyperbolic manifold $F_{[d]}$. The bulk lagrangian is
$L_b=\Lambda+F^2/(2d!)$, and the field equation for $F_{[d]}$ can
be solved by the ansatz $F_{45..d+4}=B$ with $B$ independent of
the hyperbolic coordinates. Using that there is a trapped magnetic
flux on the compact hyperbolic space, the constant $B$ is related
to the ``radion" by the equation $B=bM^{(d+4)/2}(R_h/r)^d$.
Therefore the on-shell form of the bulk lagrangian is
$L_b=M^{d+4}\lambda+M^{d+4}{b^2\over 2}({R_h\over r})^{2d}$, and
the equation \ref{eqn1} reduces to \bea
&&\lambda +{b^2 \over 2} +d(d-1)\beta^2=0, \nonumber \\
&&b^2d+d(d-1)(d+2)\beta^2=0,
\eea
where $\beta^{-1}=R_hM$. The
corresponding solution is $\lambda=-(d-1)(d-2)\beta^2/2$, and
$b^2=-(d-1)(d+2)\beta^2$, but the on-shell stress energy tensor of
the d-form is $T_{AB}={(d-1)(d-2)\beta^2\over
8}(g_{AB}-\delta^{ij}_{AB}g_{ij})$, therefore when contracted
twice with a null vector along one of the hyperbolic directions
(i.e. $l=l^0e_0+l^ie_i$), we get $T\cdot l\cdot
l=-{(d-1)(d-2)\beta^2\over 8}g_{ii}l^il^i<0$, violating the null
energy condition.

\section{FRW Brane-world and d-dimensional Hyperbolic manifold}

Let us assume now that the metric of the four dimensional
space-time is a of the form of a FRW metric, \be
ds^2=-dt^2+a(t)^2d\sigma^2\;, \ee where $d\sigma^2$ stands for the
spatial part of the metric, with curvature $k=(+1,-1,0)$. We can
repeat the calculations of the previous section with this new
metric obtaining the following set of equations:
\bea
&T_{\mu\nu}-\frac{d(d-1)M^{d+2}}{2R^2_h}G_{\mu\nu}=
(R_{\mu\nu}-{1\over 2}RG_{\mu\nu})M^{d+2},& \nonumber \\
&dG^{\mu\nu}T_{\mu\nu}-2G^{ij}T_{ij}=\frac{d(d+2)(d-1)M^{2+d}}{R_h^2},&
\eea where $R_{\mu\nu},R$ are the Ricci tensor and the Ricci
scalar of the four dimensional space-time and derivative respect
to time are wrote as dots. Then, using the usual form for the
stress energy tensor (see equation \ref{eqn2}) we get,
 \bea
&&\rho= \left[3\left({k\over a^2}+({\dot{a}\over a})^2\right)
-\frac{d(d-1)}{2R^2_h}\right]M^{2+d}, \nonumber\\
&&p=\left[ -\left({k\over a^2}+({\dot{a}\over a})^2 +
2{\ddot{a}\over a}
\right) + \frac{d(d-1)}{2R^2_h}\right]M^{2+d},\nonumber\\
&&q=\left[-3\left({k\over a^2}+({\dot{a}\over a})^2+{\ddot{a}\over
a}\right)+\frac{(d-1)(d-2)}{4R^2_h}\right]M^{2+d}. \label{eqn4}
\eea
Then, using general null vectors, we have to impose the
following system of inequalities in order to satisfy the null
energy condition:
\bea
&{k\over a^2}+({\dot{a}\over a})^2 -{\ddot{a}\over a}\geq 0,& \\
&{\ddot{a}\over a} \leq -\frac{(d-1)}{3R^2_h},&\\
&2{k\over a^2}+2({\dot{a}\over a})^2 + {\ddot{a}\over a}\geq
\frac{d(d-1)}{R^2_h},&\\
&2{k\over a^2}+2({\dot{a}\over a})^2 + {\ddot{a}\over a} \geq
\frac{(d-1)^2}{3R^2_h}.& \eea Note that if the first inequality is
saturated then the third is irrelevant since the third inequality
comes multiplied by the first inequality originally. The same is
truth for second and fourth. After some analysis we get the
following conclusions:
\begin{itemize}
\item If the first is saturated also the second has to be saturated
to get a solution, then the space-time is of the form
$AdS_4\otimes H_d$, and the curvature of $Ads_4$ is same as $H_d$.
\item If the second is saturated but not the first, the resulting
space time is not geodesically complete, as a parts of space has
to be cut off.
\item If third or fourth are saturated, there is no solution.
\item If none of the inequalities are saturated then the second inequality
rules out any physical solution as the acceleration of the
brane-world radius  $a(t)$ measure in natural units is huge,
producing incompatibilities with phenomenological data.
\end{itemize}
\vspace{.5cm}

\textit{ Although in principle equation \ref{eqn4} can be
satisfied, the matter field required will not satisfy the null
energy condition or the solutions will not be relevant in this
context.}

\section{FRW Brane-world, d-dimensional Hyperbolic manifold and n-spheres}

Let us add to the previous space-time an n-dimensional sphere of
radius $r_s$ i.e. $d(sphere)_n^2=r_s^2d\omega_{ab}dx^adx^b$ where
$d\omega_{ab}dx^adx^b$ corresponds to the metric of an n-sphere of
unit radius with volume $\Omega_n$. We have to modify our previous
definitions by: the 4-dimensional plank mass is given by
$M^2_4=M^{2+d+n}(R_h^de^\alpha)(R_s^n\Omega_n)$, we define the new
field $\psi$ by the equation $r_s=R_s
e^{\sqrt{1/n(n+2)}\psi/M_4}$. Also, we need a new conformal
rescaling on the brane metric $\bar{g}_{\mu\nu}=g_{\mu\nu}
e^{[-\sqrt{d/(d+2)}\phi/M_4-\sqrt{n/(n+2)}\psi/M_4]}$, to decouple
the fields $(\phi,\psi)$ and the reduced Einstein tensor. After
some algebra we get, \bea S_{4+d+n}=\int
dx^{d+n}{\sqrt{\gamma}\over e^\alpha}{\sqrt{\omega} \over
\Omega}S_{eff}, \eea where \bea S_{eff}=\int dx^{4}\sqrt{-g}\left[
M^2_4R(g)-{1\over 2}(\nabla\phi)^2
- {1\over 2}(\nabla\psi)^2 \right. \nonumber \\
\left. -2\sqrt{{nd\over(d+2)(n+2)}}\nabla\phi\nabla\psi
-W(\phi,\psi,g)\right], \eea and the potential is given by the
expression \bea W(\phi,\psi,g)=\frac{M_4^2}{M^{n+d+2}}{\cal
L}_{bulk}\;
e^{[-\sqrt{d/(d+2)}\phi/M_4-\sqrt{n/(n+2)}\psi/M_4]}+ \nonumber \\
(\frac{d(d-1)}{R_h^2}-\frac{n(n-1)}{R_s^2})M_4^2\;
e^{[-\sqrt{(d+2)/d}\phi/M_4-\sqrt{(n+2)/n}\psi/M_4]}. \eea where
we used that $R(\omega)=n(n-1)$. The stabilization of the two
radions translates into the following system of equations: \bea
&\partial_\phi W|_{(\phi,\psi)=0}=0\;\;\;,\;\;\;\partial^2_\phi
W|_{(\phi,\psi)=0}>0& \nonumber \\
&\partial_\psi W|_{(\phi,\psi)=0}=0\;\;\;,\;\;\;\partial^2_\psi
W|_{(\phi,\psi)=0}>0& \nonumber \\
&\left(g_{\mu\nu}W -2{\partial W \over \partial g^{\mu\nu}}
\right)|_{(\phi,\psi)=0}=0.& \eea In terms of the stress energy
tensor we get \bea &T_{\mu\nu}+{1 \over
2}(\frac{n(n-1)}{R_s^2}-\frac{d(d-1)}{R_h^2})M^{n+d+2}G_{\mu\nu}=
(R_{\mu\nu}-{1\over 2}RG_{\mu\nu})M^{n+d+2}\;,& \nonumber \\
&dG^{\mu\nu}T_{\mu\nu}-2G^{ij}T_{ij}=\frac{d(d+2)(d-1)M^{2+n+d}}{R_h^2}\;,& \nonumber \\
&dG^{\mu\nu}T_{\mu\nu}-2G^{ab}T_{ab}=-\frac{n(n+2)(d-1)M^{2+n+d}}{R_s^2}\;.&
\eea Assuming the that $T_{ab}=eR_s^2\omega_{ab}$ and equation
\ref{eqn2} we the following set of equations,
\bea &&\rho=
3\left({k\over a^2}+({\dot{a}\over a})^2\right)
+{1\over 2}\left(\frac{n(n-1)}{R^2_s}-\frac{d(d-1)}{R^2_h}\right)M^{2+d+n}, \nonumber\\
&&p=-\left({k\over a^2}+({\dot{a}\over a})^2 + 2{\ddot{a}\over
a}\right)
-{1\over 2}\left(\frac{n(n-1)}{R^2_s}-\frac{d(d-1)}{R^2_h}\right)M^{2+d+n},\nonumber\\
&&q=-3\left({k\over a^2}+({\dot{a}\over a})^2+{\ddot{a}\over
a}\right)
-{1\over 2}\left(\frac{2n(n-1)}{R^2_s}-\frac{(d-1)(d-2)}{R^2_h}\right)M^{2+d+n},\nonumber \\
&&e=-3\left({k\over a^2}+({\dot{a}\over a})^2+{\ddot{a}\over
a}\right) -{1\over
2}\left(\frac{(n-1)(n-2)}{R^2_s}-\frac{2d(d-1)}{R^2_h}\right)M^{2+d+n}.
\label{eqn5} \eea Then, the null energy condition impose the
following system of inequalities: \bea
&{k\over a^2}+({\dot{a}\over a})^2 -{\ddot{a}\over a}\geq 0,& \label{eqn6} \\
&2{k\over a^2}+2({\dot{a}\over a})^2 + {\ddot{a}\over a}\geq
{1\over 2}(-\frac{n(n-1)}{R^2_s}+\frac{d(d-1)}{R^2_h}),& \label{eqn7} \\
&{\ddot{a}\over a} \leq -{1\over 6}(\frac{n(n-1)}{R^2_s}+\frac{2(d-1)}{R^2_h}),&\label{eqn8}\\
&2{k\over a^2}+2({\dot{a}\over a})^2 + {\ddot{a}\over a} \geq
{1\over 6}(-\frac{3n(n-1)}{R^2_s}+\frac{2(d-1)^2}{R^2_h}),& \label{eqn9} \\
&{\ddot{a}\over a} \leq {1\over 6}(\frac{2(n-1)}{R^2_s}+\frac{d(d-1)}{R^2_h}),& \label{eqn10}\\
&2{k\over a^2}+2({\dot{a}\over a})^2 + {\ddot{a}\over a} \geq
{1\over 6}(-\frac{2(n-1)^2}{R^2_s}+\frac{3d(d-1)}{R^2_h}).&
\label{eqn11} \eea Note that if the first inequality is saturated
then the second is irrelevant since the second inequality comes
multiplied by the first inequality originally. The same is truth
for the third and fourth, and fifth and sixth. After some analysis
we find that:
\begin{itemize}
\item If inequality \ref{eqn6} is saturated, the solution is not useful
as the Ricci scalar is of the same order as the Hyperbolic
manifold.
\item If inequality \ref{eqn7} is saturated the resulting
space time is not geodesically complete, as a parts of space have
to be cut off.
\item If inequality \ref{eqn10} is saturated, the Ricci
scalar of the brane-world is bigger than the biggest curvature
scalar of any of the compact spaces.
\item If any of the reaming inequalities are saturated there are no solutions to the system.
\item If none of the inequalities are saturated, although in principle there could be
solutions to the system, by means of inequality \ref{eqn8} we get
back to situation in the previous section, where the acceleration
of the brane-world radius $a(t)$ measured in natural units is
huge, producing incompatibilities with phenomenological data..
\end{itemize}
Therefore the conclusion of the previous analysis is:
\vspace{.5cm}

\textit{Although in principle equation \ref{eqn5} can be
satisfied, the matter field required will not satisfy the null
energy condition or the solutions will not be relevant in this
context.}

\section{Comments on CHBS and string theory}

If you want to insist on the possibility of hyperbolic brane-world
scenarios, some of the initial assumptions have to be relaxed to
bypass the no-go theorem. Because these scenarios are thought to
be induced from string theory or M-theory, there are a couple of
generalization that can be introduced within the framework of the
different string dualities.

First let us consider the possibility of more general geometries
than the one used. Basically we can look for ten or eleven
dimensional supersymmetric solutions with hyperbolic spaces built
in. In general lets start with the action \be S_{eff}= \int dx^{D}
\sqrt{-G}(R-\sum_k{1\over 2k!}F^2_{[k]}), \ee in $D$-dimensions,
with a few k-form field strengthes $F_k$. For example, assuming we
have three fields $F_d\,,F_p\,,F_q\,,$ where each one is
proportional to the corresponding volume-form on the different
subspace where the three field strength lives, we get the
following space-time types of solutions:
\bea
AdS_d\otimes H_p\otimes S^q , \nonumber \\
AdS_d\otimes H_p\otimes T^q , \nonumber \\
AdS_d\otimes H_p\otimes H^q. \eea Where the different solutions
correspond to different ratios between the proportionality
constants of the fields $F_k$. The $AdS$ part of the solution can
not be understood as our brane-world, as its curvature is
proportional to the hyperbolic manifold curvature, giving no
useful phenomenological models. Nevertheless, what you can
certainly do is to pick up a solution of the form $AdS_5\otimes
H_d\otimes something$ and instead of thinking of it as a solution
of low energy critical string theory, use it as a solution of a
non-critical string theory, where one of the $AdS$ directions is
identify with the Liouville field. In this case taking the usual
holographic coordinates in $AdS_5$ we get a flat 4-dimensional
brane world, with a hyperbolic manifold and a holographic
direction giving information about the different energy scales
probed by the theory. This type of interpretation is along the
same line of reasoning of the confining string of Polyakov
\cite{pol1}, and also in the Verlinde et. al. constructions
\cite{ver1}. Of course in the ultraviolet limit there will be no
gravitation on the brane (only a flat brane with no gravitons
propagating on it), but at intermediate energies there will be
gravity on the brane, and generalizations to curved branes are
possible. This picture will look like a Randall-Sundrum curved
brane-world.

Another arena where to look for CHBS is within the context of type
IIA*/IIB* supergravity. These supergravity theories have been
argue to result from t-dualizing along the time-like direction. As
a result of this duality the signature of the space-time metric
can change, obtaining for example two time-like directions (see
\cite{hull}). To illustrate the mechanism that allows the
bypassing of the theorem, let us consider F-theory \cite{vafa}
(just as an example of a framework with two time-like directions).
The bulk manifold has twelve dimensions including two time-like
directions. We will be using the signature $(-+++-+++++++)$. In
this framework we are in the presence of a 4-form $F_{4}$
\cite{k1}, therefore using the ansatz
\bea
&F_{[4]}= f_{[4]}^a+f_{[4]}^b,& \nonumber \\
&f_{[4]\;4...8}^a = a\epsilon_{4...8},&  \nonumber \\
&f_{[4]\;9...11}^b = b\epsilon_{9...11},&
\eea
and decomposing the $4,5,6,7$ directions into $\tau,u,x,y$, we get
the following space-time solution, \bea &M_{(0,1,2,3)} \otimes
H_{(x,y)} \otimes S^4 \otimes
Ads_{(\tau,u)} \;\;\,\;\;\left( a = b \right),& \nonumber \\
&Ads_4\otimes H_{(x,y)}\otimes S^4 \otimes Ads_{(\tau,u)}
\;\;\,\;\;
\left( a > b \right ),& \nonumber \\
&dS_{4}\,\otimes \,H_{(x,y)}\,\otimes S^4 \otimes Ads_{(\tau,u)}
\;\;\,\;\;\left( a < b \right ).& \eea The first case has exactly
the structure we are looking for, and also note that we have the
correct $SL(2,R)$ symmetry in the part of the metric that should
account for the corresponding type IIB symmetry between the
dilaton and the axion. Thus due to the presence of an extra
time-like direction the no-go theorem is bypass.

Finally, assuming that the above frameworks bring no other
problems or conflicts with phenomenological observations, a short
discussion on the supersymmetric properties of CHBS  seems
necessary.

A question of great interest when suggesting any compactification
scheme is that of low-energy supersymmetry. For the case of
interest in this paper, we may begin with an explicit construction
of the Killing spinors of maximally symmetric spaces with negative
cosmological constant \cite{lpt1,lpr1}. For the space $H^d$ we
choose coordinates in the horospherical frame in which the metric
takes the form, \be ds^2=e^{2r}\delta_{\alpha\beta}dx^\alpha
dx^\beta+dr^2 \ . \ee In this frame, the killing spinors are given
by, \be \xi = e^{{1\over2} r\Gamma_r}\left[1+{1\over2}
x^\alpha\Gamma_\alpha(1-\Gamma_r)\right]\epsilon \ , \ee where
$\epsilon$ is an arbitrary constant spinor (the cases d=2,3 are
special and expressions can be found in ref~ \cite{fy1}).

We can see from this expression that the number of supersymmetries
of $H^d$ is equal to the number of independent spinor components.
Now, recall that the isometry group is $SO(1,d)$ and that compact
hyperbolic manifolds are obtained by quotient of $H^d$ by a
discrete subgroup $\Gamma$ of $SO(1,d)$, with no fixed points.
Whether or not any killing spinors survive this quotienting
process depends on $\Gamma$ \cite{kr1}.
\begin{enumerate}
\item If d is
even then the spinors are in an $SO(1,d-1)$ representation, and
all supersymmetries are broken.
\item If $d$ is odd, then the spinors are in an $SO(1,d)$ representation. In this case
there are several possibilities.
\begin{enumerate}
\item If $\Gamma$ is a subgroup of $SO(1,d-1)$ some killing spinors may survive, since
we can decompose the original killing spinors into Weyl spinors on
the representation of this group.
\item If $\Gamma$ is a not subgroup of $SO(1,d-1)$ then all supersymmetries are broken.
\item In the special case $d=3$ there also remain no supersymmetries.
\end{enumerate}
\end{enumerate}

\vspace{1cm} \noindent
{\bf Acknowledgments}\\

The author would like to thank the organizers of the conference {\it
Theoretical Hight Energy Physics: SUNY Utica/Rome},
where this talk was presented. This work was supported in part by
NSF grant PHY-0098747 to Syracuse University and by funds from
Syracuse University.


\end{document}